# Terahertz Magnetic and Lattice Excitations in van der Waals Ferromagnet VI$_3$

Dávid Hovančík[1]*, Dalibor Repček[2,3], Fedir Borodavka[2], Christelle Kadlec[2], Karel Carva[1], Petr Doležal[1], Marie Kratochvílová[1], Petr Kužel[2], Stanislav Kamba[2], Vladimír Sechovský[1], Jiří Pospíšil[1]

[1] Charles University, Faculty of Mathematics and Physics, Department of Condensed Matter Physics, Ke Karlovu 5, 121 16 Prague 2, Czech Republic

[2] Institute of Physics, Czech Academy of Sciences, Na Slovance 2, 182 21 Prague 8, Czech Republic

[3] Faculty of Nuclear Sciences and Physical Engineering, Czech Technical University in Prague, Břehová 7, 115 19 Prague 1, Czech Republic

Corresponding Author Email: hovancik@mag.mff.cuni.cz

**Abstract:** We use the synergy of infrared, terahertz, and Raman spectroscopies with DFT calculations to shed light on the magnetic and lattice properties of VI$_3$. The structural transition at $T_{S1}$ = 79 K is accompanied by a large splitting of polar phonon modes. Below $T_{S1}$, strong ferromagnetic fluctuations are observed. The variations of phonon frequencies at 55 K induced by magnetoelastic coupling enhanced by spin-orbit interaction indicate the proximity of long-range ferromagnetic order. Below $T_C$ = 50 K, two Raman modes simultaneously appear and show dramatic softening in the narrow interval around the temperature $T_{S2}$ of the second structural transition associated with the order-order magnetic phase transition. Below $T_{S2}$, a magnon in the THz range appears in Raman spectra. The THz magnon observed in VI$_3$ indicates the application potential of 2D van der Waals ferromagnets in ultrafast THz spintronics, which has previously been considered an exclusive domain of antiferromagnets.

TOC Graphic

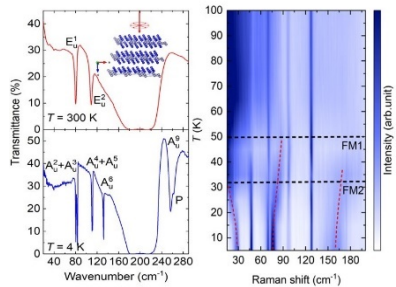

Magnetic van der Waals (vdW) materials have become a target of intense basic and applied research due to the high potential for microelectronics and spintronics. These include the transition metal ($T$) trihalides $TX_3$ ($X$ – halogen) [1] which share an X-$T$-X triple-layer structure motif consisting of a graphene-like honeycomb network of $T$ atoms in the $ab$ plane interposed between two layers of X atoms. The triple-layers are stacked on top of each other and the neighboring halogen atom layers in the sequence …. X-$T$-X-X-$T$-X … are separated by a vdW gap.

CrI$_3$, the most intensively studied $TX_3$ compound, is an Ising-type ferromagnetic (FM) semiconductor with $T_C$ = 61 K [2]. VI$_3$ adopts a trigonal structure ($R\bar{3}$ space group) at room temperature (see Figure S1, S2 and S3) and then, upon cooling, it undergoes a first-order transition to the monoclinic phase at $T_{S1}$ = 79 K [3-5] and becomes FM at $T_C$ = 50 K (see Figure S4) while keeping the monoclinic symmetry. The transition from the monoclinic to a triclinic structure ($P\bar{1}$ space group) at $T_{S2}$ = 32 K is associated with the transformation from one FM structure to another [6-9]. These structural phase transitions are associated with distortions of the V honeycomb networks and minute changes in the V-V distances [8, 9].

The measured ordered V magnetic moment in VI$_3$ (~ 1.24 $\mu_B$ at 4 K) [9] (see Figure S4) is significantly smaller than the expected spin-only moment of a V$^{3+}$ ion (2 $\mu_B$) and tilted from the V honeycomb network's normal by ~ 40° [10, 11]. When decreasing temperature from 50 K to 30 K the in-plane moment component gradually rotates within the *ab*-plane [11] - in total by 90° [9]. The complex magneto-crystalline anisotropy together with the reduced V$^{3+}$ magnetic moment can be attributed to the breaking of the three-fold symmetry in the monoclinic and triclinic phases, to a large orbital moment of V$^{3+}$ ion, and spin-orbit coupling in both V$^{3+}$ and I$^-$ ions [12].

Similar to the CrI$_3$ case [2], a closer inspection of crystallographic and magnetic data collected on VI$_3$ provides indications of strong magneto-elastic coupling (MEC) as a key underlying mechanism of the observed unusual phenomena. In particular, the crystallographic phase transition at $T_{S1}$ induces a change of magnetization in the paramagnetic state far above $T_C$ [3, 4, 6], and also the transition temperature $T_{S1}$ can be controlled by the applied magnetic field [7]. The structural phase transition observed at 32 K and the aforementioned magnetic phase transition are interconnected via a strong MEC.

Raman scattering has proven to be an effective tool for investigating the magnetic and lattice degrees of freedom in 2D magnets [13]. Although it directly probes the optical phonons, the phenomena such as MEC, zone-folding, or magneto-optical Raman effect [14, 15] provide additional information about magnetism [16-18]. The Raman study by Lyu et al. [19] performed on exfoliated flakes of VI$_3$ revealed the presence of a spin-wave excitation (magnon) and a two-magnon mode in the FM phase. Likewise, the signatures of FM order were spotted using polarization-resolved Raman scattering. Furthermore, the large quasi-elastic (QES) scattering around the temperature of the structural transition $T_{S1}$ = 79 K has been identified. None of those features has yet been detected in the bulk crystal. Infrared (IR) spectroscopy data complementary to Raman scattering data are also lacking [20]. In the case of VI$_3$, such a lack of information caused a misinterpretation of the crystallographic phases and phase transitions [19].

In this work, we used the complementarity of IR, terahertz (THz), and Raman spectroscopies to study the signatures of structural and magnetic phases in the bulk VI$_3$ single crystal at temperatures down to 4 K. This approach has allowed us to obtain a complete picture of the phonon modes in VI$_3$, which can be compared to first principles predictions. We found that at $T_{S1}$ = 79 K doubly degenerate IR active modes split and two new modes appear due to the structure symmetry lowering. At $T_C$, clear signs of MEC in both Raman and IR spectra together with the modification of Raman polarization selection rules for $A_g$ modes reflect the formation of the FM phase. In Raman spectra, we have identified a high-frequency acoustic magnon emerging below $T_{S2}$. The frequencies of the observed magnon are three orders higher than those in conventional ferromagnets. In addition, two other Raman active modes associated with ferromagnetism in VI$_3$ simultaneously emerge below $T_C$. They dramatically change their frequency in the neighborhood of $T_{S2}$ indicating unusual interplay of low-temperature magnetic and structural phase transitions.

The room-temperature IR transmittance spectrum of VI$_3$ shown in Figure 1a reveals two phonon modes located at 80 and 109 cm$^{-1}$. No transmittance is observed in the frequency range of 180-220 cm$^{-1}$, which is typical for the reststralen band [21] with almost 100 % reflectivity. The edges of this band should correspond to the frequencies of the transverse and longitudinal phonon [22]. Using the results of our first-principles calculations for the $R\bar{3}$ rhombohedral phase [4, 7, 23], we assigned the symmetries of the phonons seen in Figure 1a.. The modes visible at 300 K were identified as doubly degenerate $E_u$ modes oscillating in-plane. $E_u^1$ and $E_u^2$ show clear splitting upon cooling between 80 and 77.5 K (passing $T_{S1}$). Furthermore, the other three modes appear at 131, 256, and 262 cm$^{-1}$ just below $T_{S1}$ as seen in Figure 1c. The lifting of the phonon mode degeneracy reflects the crystal symmetry reduction due to the rhombohedral to monoclinic phase transition [3, 4, 7, 8, 23, 24]. The THz spectra measured at several temperatures (see Figure S5) confirmed the splitting of the $E_u^1$ mode at $T_{S1}$.

Given the point groups corresponding to the $R\bar{3}$ ($C_{3i}$ point group), $C2/m$ ($C_{2h}$ point group) and $P\bar{1}$ ($C_i$ point group) space groups and the Wyckoff positions of V and I atoms [5, 7] we can consider the factor group analysis for zone-center optical phonons (see Table 1). The $C_{3i}$ point group has 6 IR active modes ($3A_u + 3E_u$) and 8 Raman active modes ($4A_g + 4E_g$), the $C_{2h}$ point group 9 IR active modes ($5A_u + 4B_u$) and 12 Raman active modes ($6A_g + 6B_g$), and the $C_i$ point group 9 IR active $A_u$ modes and 12 Raman active $A_g$ modes.

**Table 1: The irreducible representations of the optical phonons.** The irreducible representations are given for $D_{3d}$, $C_{3i}$, $C_{2h}$, and $C_i$ point group [25]

| Point group | Irreducible representations for the optical modes |
|---|---|
| $D_{3d}$ | $\Gamma_{optical} = 2A_{1g} + 2A_{2g} + 4E_g + A_{1u} + 2A_{2u} + 3E_u$ |
| $C_{3i}$ ($R\bar{3}$) | $\Gamma_{optical} = 4A_g + 4E_g + 3A_u + 3E_u$ |
| $C_{2h}$ ($C2/m$) | $\Gamma_{optical} = 6A_g + 6B_g + 4A_u + 5B_u$ |
| $C_i$ ($P\bar{1}$) | $\Gamma_{optical} = 12A_g + 9A_u$ |

All doubly degenerate $E_u$ and $E_g$ phonon modes should split up at $T_{S1}$. The splitting of $E_{2u}$ mode is quite weak just below $T_{S1}$ such that only a shoulder (asymmetry) of the peak is observed; the splitting can be experimentally resolved below ~ 55 K deeply in the monoclinic phase (see Figure 1c). This result can be understood by the subtle structural change at $T_{S1}$ [7] leaving certain two-dimensional modes almost "intact". In other words, the splitting of the $E_u^2$ mode above 55 K is below the resolution of the spectrometer making it appear as a single peak.

Since the triple-layer of VI$_3$ has the $D_{3d}$ point symmetry for both bulk phases with $C_{3i}$ and $C_{2h}$ point symmetry (no relation group-subgroup) [26], we can find the relations between the irreps of the normal modes (see Table S1). For simplicity, we will label the phonon modes below $T_{S1}$ using the irreps of the $C_i$ point group. The splitting can be quantified by DFT calculations for the low-temperature structure (see Table 2.). The monoclinic symmetry below $T_{S1}$ [3, 7] permits distortions of the original V honeycomb layers that break the 3-fold rotation symmetry, namely dimerization or anti-dimerization [27]. Son et al. [3] have found evidence of anti-dimerization, while another work argues for dimerization [8]. Our previous computational study has shown that a significant perturbation of the 3-fold rotation symmetry as introduced by anti-dimerization is needed to explain the unusual angular dependence of magnetic anisotropy energy in this system [12]. We found out that the two new modes $A_u^6$ and $A_u^9$ observed below $T_{S1}$ at 131 and 256 cm$^{-1}$ (see Figure 1b.) correspond to the $A_u^2$ and $A_u^3$ modes (rhombohedral notation) that are not seen in the rhombohedral phase due to the sample orientation (see inset in Figure 1a.). The rapid increase in the activity of $A_u^6$ and $A_u^9$ mode can be explained by the first order phase transition at $T_{S1}$ [28]. An additional peak $P$ at 261 cm$^{-1}$, which also shows up below $T_{S1}$, does not correspond to any phonon mode in our calculated spectra (see Table 2.). It might possibly come from the splitting of the $E_u^3$ mode. However, such a large splitting is inconsistent with the subtle structural change at $T_{S1}$.

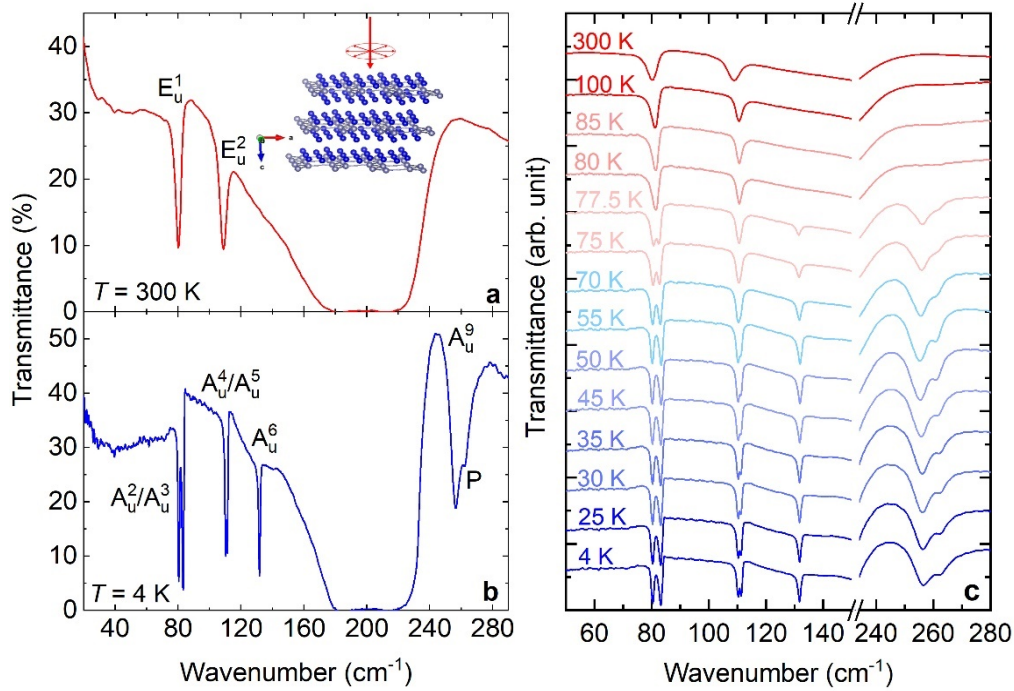

**Figure 1: IR transmittance spectra of VI$_3$ single crystal.** a) Assignment of the modes at 300 K and b) 4 K. c) Whole temperature dependence. The inset in 1. a) shows the orientation of the layers with respect to the incoming unpolarized beam. The crystal had the thickness of ~30 μm.

By fitting the spectral line shape of the phonon modes with a Lorentzian function we obtained the temperature-dependence of the mode frequencies shown in Figure 2. We noticed the mirrored splitting trend of $E_u^1$ and $E_u^2$ "daughter" modes $A_u^2+A_u^3$ and $A_u^4+A_u^5$ in the monoclinic phase. The observed splitting of the $E_u^1$ mode is rather consistent with the anti-dimerization picture which indicates that this distortion variant is present. The above-mentioned "daughter" modes and also the $A_u^6$ mode reverse the softening/hardening tendency at 55 K. This happens 5 K above $T_C$ where short-range ferromagnetic order is expected in a 2D ferromagnet, which due to strong MEC may lead to slight shifts of phonon frequencies. In this context, it is worth noticing that VI$_3$ thin films [29], monolayer of VI$_3$ [30] and also partly degraded surface layers of a bulk crystal [31, 32] exhibit a higher value of $T_C$ than bulk crystals. The temperature dependence of $A_u^2$, $A_u^3$, $A_u^4$, and $A_u^5$ phonons reverse the trend again near 30 K, although less significantly. This could be considered as a signature of the second structural transformation from the monoclinic to triclinic phase [7] conjugated with the change of FM order.

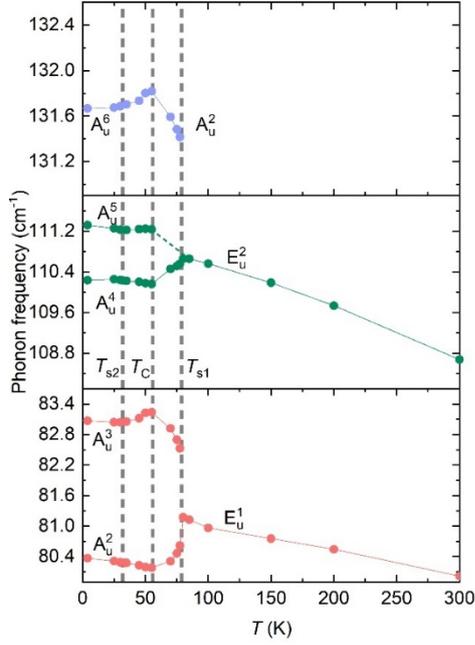

**Figure 2: Temperature dependence of the phonon's frequencies.** For the three selected IR active modes $E_u^1$, $E_u^2$ and $A_u^2$ (silent in the rhombohedral phase). The solid vertical lines indicate the temperatures of high-temperature structural transition $T_{S1}$, ferromagnetic transition $T_c$ and low-temperature structural transition $T_{S2}$.

**Table 2: Calculated Γ-point optical phonon frequencies.** For $R\bar{3}$ ($C_{3i}$) space group, monoclinic variant ($C_{2h}$) and triclinic dimerized ($C_i$) and anti-dimerized ($C_i$) version. The assignment to irreducible representations of $C_{3i}$ and $C_i$ point group is given.

| | *Rhombohedral* | | *Monoclinic* | *Triclinic* | | | |
|---|---|---|---|---|---|---|---|
| Phonon | $R\bar{3}$ ($C_{3i}$) | Experiment (300 K) | ($C_{2h}$) | ($C_i$) Anti-dimerized | | ($C_i$) dimerized | Experiment (4 K) |
| | Irrep. | Freq (cm$^{-1}$) | Freq (cm$^{-1}$) | Irrep. | | Freq (cm$^{-1}$) | |
| 1 | $E_g^1$ | 47.1 | 47.9 | 44.3 | $A_g^1$ | 38.0 | 41.0 | 47.9 |
| 2 | $E_g^1$ | 47.1 | 47.9 | 48.7 | $A_g^2$ | 49.4 | 49.5 | 47.9 |
| 3 | $A_u^1$ | *70.8* | - | 66.1 | $A_u^1$ | 70.8 | *68.1* | - |
| 4 | $A_g^1$ | 74.8 | 70.3 | 78.4 | $A_g^3$ | 77.4 | 77.9 | 69.1 |
| 5 | $E_u^1$ | 83.3 | 80.5 | 81.1 | $A_u^2$ | 82.1 | 83.6 | 80.2 |
| 6 | $E_u^1$ | 83.3 | 80.5 | 82.6 | $A_u^3$ | 84.5 | 84.1 | 83.1 |
| 7 | $A_g^2$ | 86.6 | - | 87.2 | $A_g^4$ | 88.1 | 88.1 | - |
| 8 | $E_g^2$ | 91.1 | - | 88.9 | $A_g^5$ | 92.3 | 92.6 | - |
| 9 | $E_g^2$ | 91.1 | - | 93.5 | $A_g^6$ | 96.9 | 93.9 | - |
| 10 | $E_g^3$ | 98.9 | 96.4 | 95.4 | $A_g^7$ | 101.4 | 95.7 | 96.4 |
| 11 | $E_g^3$ | 98.9 | 96.4 | 100.5 | $A_g^8$ | 112.3 | 100.5 | 100.2 |
| 12 | $E_u^2$ | 114.3 | 108.9 | 110.3 | $A_u^4$ | 114.9 | 112.7 | 110.3 |
| 13 | $E_u^2$ | 114.3 | 108.9 | 111.4 | $A_u^5$ | 120.3 | 114.7 | 111.2 |

| 14 | $A_g^3$ | 127.0 | 126.2 | 126.6 | $A_g^9$ | 130.0 | 128.4 | 127.4 |
| 15 | $A_u^2$ | 131.4 | - | 128.2 | $A_u^6$ | 132.4 | 130.4 | 131.7 |
| 16 | $A_g^4$ | 197.9 | - | 178.0 | $A_g^{10}$ | 186.1 | 185.4 | - |
| 17 | $E_u^3$ | 226.4 | - | 216.8 | $A_u^7$ | 222.0 | 222.3 | - |
| 18 | $E_u^3$ | 226.4 | - | 220.2 | $A_u^8$ | 227.3 | 224.8 | - |
| 19 | $E_g^4$ | 232.8 | - | 225.4 | $A_g^{11}$ | 231.2 | 227.2 | - |
| 20 | $E_g^4$ | 232.8 | - | 229.3 | $A_g^{12}$ | 238.7 | 241.1 | - |
| 21 | $A_u^3$ | 255.0 | - | 244.1 | $A_u^9$ | 247.6 | 250.0 | 256.7 |

Figure 3 shows Raman spectra of bulk VI$_3$ measured in the back-scattering geometry at $T$ = 300 K and 4 K. Four Raman active modes are identified at 300 K with frequencies of 48, 70, 96, and 126 cm$^{-1}$ - see Figure 3a. The Raman intensities from the parallel (XX) and crossed (XY) polarization configuration spectra give us information about the symmetry of the Raman tensor for a given mode. The Raman intensity is proportional to $|e_s^\dagger \overleftrightarrow{R} e_i|^2$, where $\overleftrightarrow{R}$ is the Raman polarizability tensor, and $e_s$ and $e_i$ are the polarization vectors of the scattered and incident light (off-resonant scattering) [33]. By using this definition and the Raman tensors for the irreps of $C_{3i}$ point group with the form:

$$R(A_g) = \begin{pmatrix} a & & \\ & a & \\ & & b \end{pmatrix} \quad (2)$$

$$R(^1E_g) = \begin{pmatrix} c & d & e \\ d & -c & f \\ e & f & \end{pmatrix} \quad (3)$$

$$R(^2E_g) = \begin{pmatrix} d & -c & -f \\ -c & -d & e \\ -f & e & \end{pmatrix} \quad (4)$$

we found two modes with $A_g$ symmetry and two with $E_g$ symmetry, similarly to the previous studies [24, 34]. The set of *gerade* phonon frequencies calculated for $R\bar{3}$ using the DFT that are given in Table 2 show very good agreement with the experimental results. We note that this result contradicts the Raman study of Djurdic et al. [24] claiming better agreement with the $P\bar{3}1c$ space group ($D_{3d}$ point group) rather than with $R\bar{3}$.

The Raman spectrum at 4 K shown in Figure 3b corresponds to the triclinic FM phase (the phonons are labeled using $C_i$ point group irreps) and contains three additional modes at 28.7, 76, and 160 cm$^{-1}$ labeled as $\Omega_1$, $\Omega_2$, $\Omega_3$. A sharp $\Omega_1$ and a broad $\Omega_3$ resembles the modes observed below $T_C$ by Lyu et al. [19]. They were ascribed to the zone-center acoustic magnon (two magnetic branches expected for two V atoms per unit cell) and two-magnon mode associated with the Van Hove singularity of the magnon density of states (DOS) corresponding presumably to the zone-boundary magnons. Besides, a small but clear splitting of the $E_g^3$ phonon into two modes $A_g^7$ and $A_g^8$ is present according to the opposite polarization selection rules.

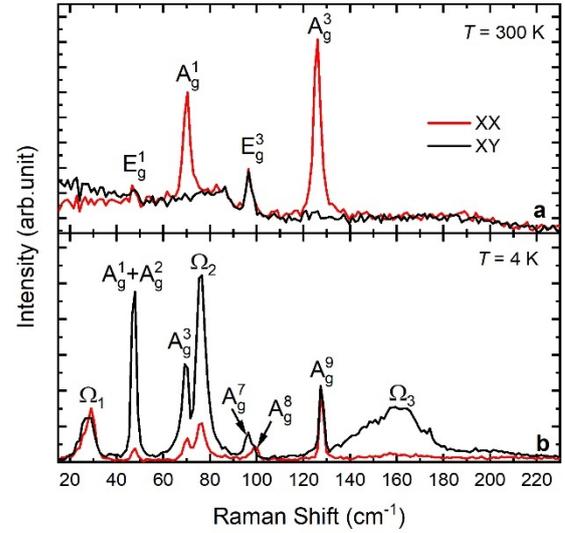

**Figure 3: Polarization-resolved Raman spectra of VI$_3$. a,** taken at 300 K in the rhombohedral phase. **b,** taken at 4 K in the ferromagnetic triclinic phase. The Raman active phonons are named after the irreps of the given point group.

The temperature evolution of the Raman spectra from 300 K to 4 K is given in Figure 4a. No lifting of the $E_g$ modes degeneracy is found at $T_{S1}$ within the experimental energy resolution (~ 1 cm$^{-1}$), similarly to what was found for RuCl$_3$ [35]. Our DFT calculations show a relatively large splitting of the $E_g^1$ and $E_g^3$ modes ($\Delta\omega$ ~10 cm$^{-1}$ and ~ 20 cm$^{-1}$) when anti-dimerization is included while splitting due to dimerization is significantly smaller. On the other hand, this splitting is rather asymmetric and one of the peaks practically retains the position of its high symmetry parent. The intensity of the other component could be below the experimental sensitivity. Below $T_C$, two new features emerge in Raman spectra, a sharp $\Omega_2$ mode accompanied by the formation of a broad $\Omega_3$, which develop and soften with further decreasing temperature. Below 35 K, an $\Omega_1$ mode emerges just above our energy cut-off (12 cm$^{-1}$) and exhibits hardening with decreasing temperature down to 4 K. The hardening/softening trends of $\Omega_1$, $\Omega_2$ and $\Omega_3$ modes with decreasing temperature are much more significant compared to the Γ-point optical phonons. It is well documented in Figure 4b showing the temperature dependence of the frequencies of the Raman modes.

While the phonon frequency changes are within 3 cm$^{-1}$, the frequency shifts of $\Omega_1$, $\Omega_2$ and $\Omega_3$ approach 10, 11 and 18 cm$^{-1}$, respectively, in the temperature range between 45 K and 4 K (for $\Omega_1$ between 25 K and 4 K). Such a strong temperature dependence of frequencies indicates that these three modes are connected with the evolution of magnetic structure [9].

The $\Omega_1$ and $\Omega_2$ modes shift to higher frequencies with an increasing magnetic field, directly proving their magnetic origin. This aspect is presented here through the courtesy of Dr. M. Veis et al. as a result of their recent measurements of Raman spectra on VI$_3$ flakes (from the same batch as our samples) in the magnetic field [36]. The evolution of $\Omega_1$ with decreasing temperature qualitatively agrees with that observed for the acoustic magnon in [19]. This acoustic magnon originates from the out-of-plane magneto-crystalline anisotropy (MCA) creating a gap at the Γ-point in the magnon dispersion which is proportional to the MCA energy [19, 37]. The assignment of the $\Omega_1$ mode to a magnon is further supported by recent inelastic neutron scattering measurements that observed a Γ-point magnetic excitation with a gap of ~ 4 meV (~32 cm$^{-1}$) at 2 K [38], which almost coincides in frequency with $\Omega_1$ (~29 cm$^{-1}$, ~0.8 THz). Because of differences in frequency of $\Omega_1$ obtained by different Raman experiments, we performed additional testing measurements on different samples, which revealed that the $\Omega_1$ frequency may get lower, ~ 20 – 25 cm$^{-1}$, and that this depends on the laser spot location in the sample (see Figure S6). These values are closer to the acoustic magnon energy of ~ 18.5 cm$^{-1}$ at 1.7 K reported in [19]. In any case, the observed ferromagnetic magnon mode $\Omega_1$ is three orders of magnitude higher in energy than those observed in conventional ferromagnets [39].

As for the $\Omega_2$ mode one may consider assigning it to an optical magnon originating from the upper branch in the $\Omega_1$ magnon dispersion or to a second acoustic magnon. The existence of two acoustic magnon branches in VI$_3$ with two different gaps at the Γ-point was observed by inelastic neutron scattering [38] and explained by associating them with two coexisting VI$_3$ structure domains hosting different V$^{3+}$ orbital ground states due to oppositely distorted octahedral V$^{3+}$ environments. This picture is consistent with the two inequivalent V moment sites observed by NMR and supported by muon experiment[40], one of them with an ordered magnetic moment existing only below $T_{S2}$ and the other up to $T_C$ [6]. The $\Omega_2$ mode frequency at 4 K (~76 cm$^{-1}$) is, however, 26 % higher compared to the upper Γ-point magnetic excitation of ~ 7 meV (~56 cm$^{-1}$) seen by neutrons. The broad $\Omega_3$ mode does not show any magnetic-field dependence [19]. Its large width of 27 cm$^{-1}$ at 4 K indicates a second-order scattering [41]. The analogous temperature dependence of the frequencies of $\Omega_2$ and $\Omega_3$, where the frequency of $\Omega_3$ is roughly twice as high as that of $\Omega_2$ (see Figure 4b), suggests that they might originate from the same branch.

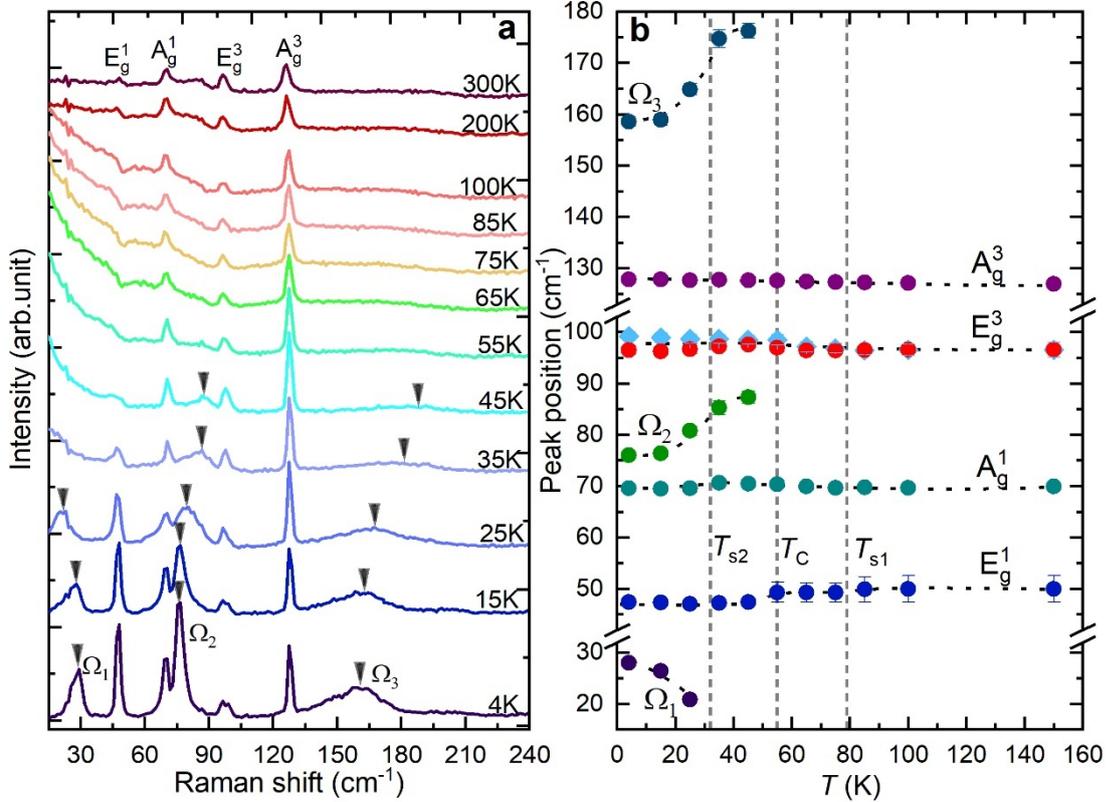

**Figure 4: Temperature evolution of the unpolarized Raman spectra. a,** temperature-dependent unpolarized Raman spectra taken from 300 K to 4 K in zero magnetic field. The $\Omega_1$ mode emerges only below $T_{S2}$ whereas, $\Omega_2$ and $\Omega_3$ survive up to $T_C$, (marked by arrows). **b,** peaks positions in frequency as a function of temperature obtained from the spectra presented in **a,**. For $E_g^3$ mode the peak position measured in XX (blue diamonds) and XY (red circles) polarization configuration is given as the mode exhibits splitting under different polarization configurations. The peaks were fitted with the pseudo-Voigt function except for the $E_g^1$ mode where we used a Fano resonance shape function.

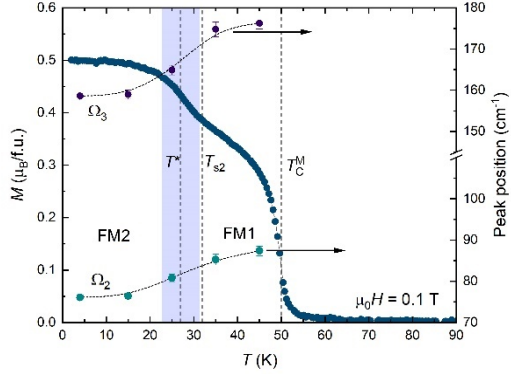

**Figure 5: The temperature dependence of frequencies of $\Omega_2$ and $\Omega_3$ mode.** The magnetization as a function of temperature measured after zero-field cooling with the magnetic field applied within the *ab*-plane is shown together with the temperature dependence of $\Omega_2$ and $\Omega_3$ mode's frequencies. The plot should emphasize the correlation of the Raman modes with the in-plane magnetic moment rotation happening around the temperature of the second structural transition.

To investigate the connection of $\Omega_2$ and $\Omega_3$ modes with magnetism, we compare the temperature dependence of frequencies of these two modes with the in-plane magnetization - see Figure 5. We note that both modes dramatically soften around $T^*$ (just below $T_{S2}$) where the in-plane spin reorientation, causing a change of the magnetization, has been observed [9]. This cross-correlation further supports that $\Omega_2$ and $\Omega_3$ are of magnetic origin.

Next, we analyze the temperature dependence of the polarization selection rules of different modes in the Raman spectra seen in Figure 6a. The Raman tensors for the monoclinic phase with the point group symmetry $C_{2h}$ are derived as:

$$R(A_g) = \begin{pmatrix} a & & d \\ & c & \\ d & & b \end{pmatrix} \quad (5)$$

$$R(B_g) = \begin{pmatrix} & e & \\ e & & f \\ & f & \end{pmatrix} \quad (6)$$

Since in our experimental setup (backscattering geometry with the polarization of the incident and scattered light parallel to the V honeycomb planes) we probe only the elements of the top left two-by-two submatrix of Equation (5), the $A_g$ modes are forbidden in the XY channel below $T_{S1}$. This is consistent with our results where the $A_g^3$ ($A_g^1$ in rhombohedral phase) and the $A_g^9$ ($A_g^3$ in the monoclinic phase) modes remain silent in the XY polarization configuration. However, both modes start to appear in the XY channel at $T \approx 55$ K. Such polarization change indicates a modification of selection rules due to the FM order breaking time-reversal symmetry which was also observed in CrI$_3$ [14]. The temperature-dependent polarization intensity ratio $I_{xy}/I_{xx}$ of the dominant $A_g^9$ mode shown in Figure 6c demonstrate this effect. A relevant explanation can be obtained from the group theory analysis. Given the V$^{3+}$ magnetic moment alignment above $T_{S2}$ [9], the magnetic point group of bulk VI$_3$ is likely $C_{2h}(C_i) = \{e, RC_2, i, R\sigma_h\}$, where $R$ is the time-reversal operator. In the FM state, the Raman tensor for the $A_g$ optical phonon has the same form as that for the optical magnon because they have identical symmetry [42]. Therefore, the Raman tensor for the $A_g$ mode given by the co-representation of the $C_{2h}(C_i)$ magnetic point group [14, 43] has the form

$$R_{FM}(A_g) = \begin{pmatrix} a & ih & d \\ ig & c & ik \\ d & ij & b \end{pmatrix} \quad (7)$$

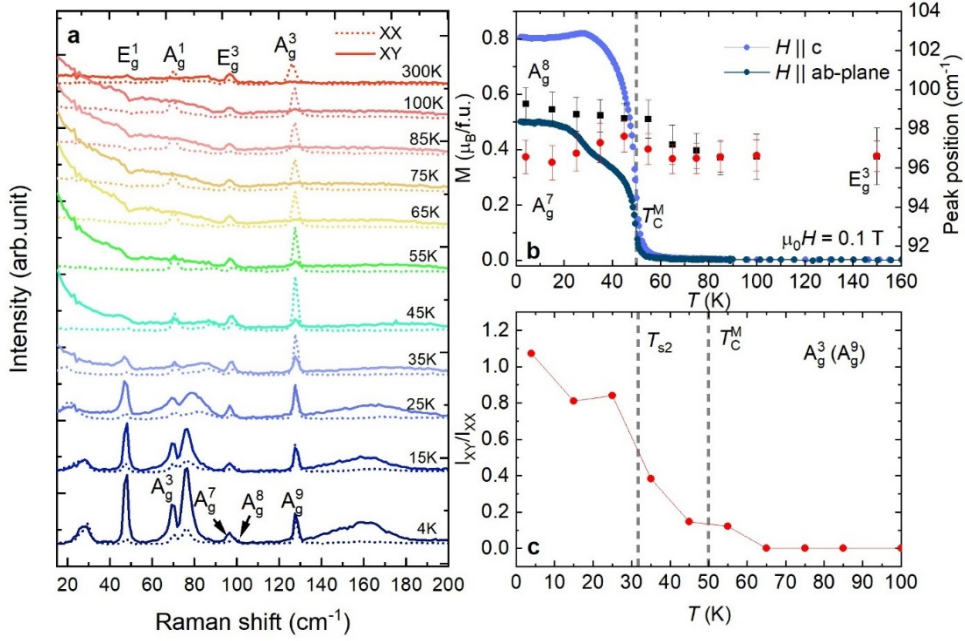

**Figure 6: Temperature dependence of the Polarized Raman spectra. a,** Polarized Raman spectra of bulk VI$_3$ as a function of temperature. **b,** The temperature dependence of the frequency of $E_g^3$ phonon mode in XX (black squares) and XY (red circles) polarization configuration and magnetization of VI$_3$. **c,** The polarization intensity ratio $I_{xy}/I_{xx}$ for $A_g^3$ mode as a function of temperature.

The imaginary off-diagonal terms in Equation (7) can explain the non-zero Raman scattering intensity in the XY polarization configuration present between $T_C$ and $T_{S2}$.

A closer inspection of the Raman spectra in Figure 6a reveals the splitting of the $E_g^3$ phonon mode as the peak position is slightly different under XX and XY polarization configurations. Experimentally, we observe almost no separation of the two modes ($A_g^7 + A_g^8$) in the paramagnetic phase measured in both polarization configurations. Nevertheless, in the FM phase, the vibrational frequencies shift oppositely with decreasing temperature as shown in Figure 6b. We tentatively attribute this effect to a MEC. A decrease of one phonon frequency on cooling below $T_C$ indicates ferromagnetic interactions between ions involved in this phonon vector [44,45] as was observed in NiPS$_3$ [18]. Higher-resolution Raman spectroscopy is, however, necessary to accurately determine the mode-splitting temperature.

It is worth making a few remarks about the polarization properties of the $\Omega_1$ magnon. The well-known theory of Fleury and London [41] predicts the purely antisymmetric (present only in the XY configuration) Raman scattering from the first-order magnetic excitation regardless of crystal orientation. This theoretical prediction successfully explains a variety of ferromagnetic and antiferromagnetic materials with Raman active magnons [41, 46, 47]. However, the theory uses the approximation of a zero or quenched angular momentum ground state L = 0 [41] which appears not to be true for VI$_3$ [12, 48]. The recent work of McCreary et al. [49] on vdW FePS$_3$ pointed out the deviation from the purely antisymmetric response of a magnon and ascribed it to the non-zero L. In our case, we observe the $\Omega_1$ magnon in both XX and XY polarization configurations which is in stark contrast with the Fleury and London theory. This anomaly further corroborates the suggested picture of a high orbital momentum on V$^{3+}$ ion [12, 48, 50].

Furthermore, we notice a large contribution from the quasi-elastic scattering (QES) centered around the Rayleigh line which is present mainly in the XY polarization configuration shown in Figure 6a. The QES is usually detected in highly frustrated and low-dimensional magnetic systems. The fluctuations of magnetic energy density are believed to be the source of this phenomenon [16, 35, 51, 52]. Since the VI$_3$ has low frustration index [53], we attribute this effect to the quasi-2D nature of the magnetism. The intensity of the QES increases upon cooling and reaches its maximum at around 80 K and 65K for the XX and XY polarization configurations, respectively. Then it steadily decreases turning into the $\Omega_1$ acoustic magnon at around 35 K. Following the theory of Riter [54] and Halley [55] the Raman response for Stokes scattering defined as $\chi''(\omega) = I(\omega)/[n(\omega) - 1]$ where $n(\omega)$ is the Bose-Einstein factor, is given as [56, 57]

$$\frac{\chi''(\omega)}{\omega} \propto C_m T \frac{Dk^2}{\omega^2 + (Dk^2)^2} \quad (8)$$

where $C_m$ is magnetic specific heat and $D$ is thermal diffusion constant. To obtain the spectral weight of QES we fit the normalized Raman response $\chi''(\omega)/\omega$ data with a Lorentzian function and integrate within the frequency range of 12-100 cm$^{-1}$ (see Figure S7). The temperature dependence of the spectral weight of QES for both XX and XY polarization configurations is given in Figure 7a. We can note that the QES spectral weight peaks around $T_{S1}$ for XX polarization configuration, whereas for XY the dominant peak shows its maximum around 65 K. These results provide microscopic evidence of the presence of FM fluctuations as a precursor of long-range FM order, far above $T_C$ in the paramagnetic state. Interestingly, the FM fluctuations survive even below $T_C$ as displayed by the non-negligible QES spectral weight contribution probably coming from the still fluctuating V moment.

Another interesting feature we notice is the temperature evolution of the prominent spectral line shape of the $E_g^1$ (corresponds to $A_g^1+A_g^2$ below $T_{s1}$) phonon mode. As Figure 4a and 6a show, this mode exhibits clear characteristics of a Fano resonance; slight asymmetry at high temperatures followed by the mode appearing as a dip in the spectrum superposed on QES which turns into an asymmetric peak at around 55 K, and finally, the asymmetry vanishes at the low-temperature limit. The Fano resonance requires quantum interference between a discrete excitation and a continuum [58] and the resonance line shape is described by the form

$$I(\omega) = I_0 \frac{[1 + (\omega - \omega_0)/q\Gamma]^2}{1 + [(\omega - \omega_0)/\Gamma]^2} \quad (9)$$

where $\omega_0$ and $\Gamma$ are the bare frequency and width of the phonon, and $q$ is the asymmetry parameter, where $1/|q|$ correlates with the strength of the coupling. In this case, the discrete excitation and the continuum correspond to the $E_g^1$ phonon and magnetic continuum states, respectively. Therefore, the $1/|q|$ parameter is proportional to the spin-phonon coupling strength [59]. With the subtraction of the QES we fitted the $E_g^1$ phonon with Equation (9) (see Figure S8) with the $1/|q|$ plotted in Figure 7b. The $1/|q|$ parameter shows a maximum in the temperature range of the first structural transition where the QES is relatively strong and the spectral weight of QES spans a wide frequency range. This suggests a tight coupling between the phonon mode and the FM fluctuations which results in the anti-resonance. Interestingly, the striking magnetic field dependence of the structural transition at $T_{s1}$ observed by Dolezal et al. [7] could be potentially connected to this effect. At the low-temperature limit, where the QES is significantly suppressed by FM order, we observe a negligible value of $1/|q|$, which is a characteristic of regular resonance dominated by phonon scattering. We note, that a similar evolution of the $E_g^1$ phonon mode with temperature was reported for VI$_3$ thin films [19].

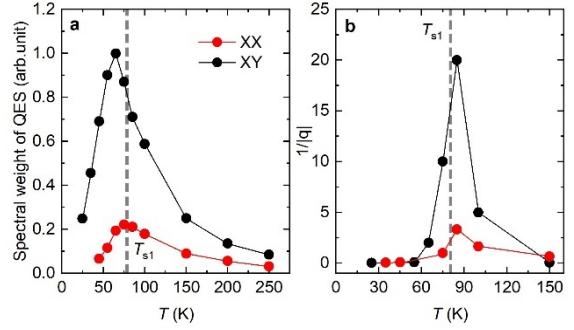

**Figure 7: Temperature dependence of the Spectral weight of the QES and the Fano asymmetry parameter. a,** the integrated intensities of the QES plotted as a function of temperature for both polarization configuration. Both dependences show maximum around the temperature of the first structural transition $T_{S1}$ (~ 80 K and 65 K for XX and XY channel, respectively). **b,** the inverse Fano asymmetry parameter for $E_g^1$ phonon mode as a function of temperature for both polarization configurations peaks at around $T_{s1}$ which indicates strong coupling between the phonon mode and the magnetic continuum.

In summary, we have studied the lattice and magnetic properties of single crystals of the vdW compound VI$_3$ at temperatures down to 4 K by combining infrared, terahertz, and Raman spectroscopy and employing relevant DFT calculations of phonon modes. The observed results confirm and complement the state of understanding of the evolution of crystal structure and magnetic phases with varying temperatures. The IR and THz spectra revealed splitting of phonon modes upon cooling of the crystal through the structural transition at $T_{S1}$ which confirms the symmetry lowering due to the change of the trigonal phase to monoclinic. In addition, the large enhancement of the QES between $T_{S1}$ and $T_C$ detected by the Raman scattering reflects the presence of strong FM fluctuations at temperatures far above $T_C$ which are tightly coupled to at least one phonon mode. The onset of the long-range FM order is apparent from observed variations of phonon frequencies at $T_C$ via MEC enhanced by strong spin-orbit interaction. Moreover, the modification of the Raman polarization selection rules for optical phonons at $T_C$ unambiguously demonstrates that Raman spectroscopy can be used as a simple probe of magnetic phases for $VI_3$ and potentially for other vdW compounds.

The Raman spectra revealed a high-frequency acoustic magnon in the THz range emerging below the critical temperature $T_{S2}$ of the second structural transition accompanied by order-order magnetic phase transition. In addition, two new Raman active modes are observed below $T_C$. They show a dramatic softening in the narrow temperature interval around $T_{S2}$ in which the order-order magnetic phase transition involving rotation of in-plane magnetic moment has been reported [9]. The THz range of magnon frequency in $VI_3$ is three orders of magnitude higher than typical GHz magnon frequencies in conventional ferromagnets. This suggests that some 2D van der Waals ferromagnets can be useful in ultrafast THz spintronic applications, which were believed to be exclusively the domain of antiferromagnets.

Methods

*Sample Fabrication*. The $VI_3$ single crystals have been grown through the chemical vapor transport method directly from a stoichiometric (1:3) ratio of 2N5 purity V powder (Goodfellow) and 5N ultra dried $I_2$ lumps (Thermofisher) in a quartz ampoule evacuated down to $10^{-7}$ bar. Preparation of elements as well as the filing of the quartz ampoule was performed in a glove box under Ar inert atmosphere. The growth process was realized in a two-zone gradient furnace. Two processes were tested. In the first case, the procedure reported by Tian et al. [5] was followed. The starting material was gradually heated up to 650°C for 4 days. Then a gradient of 650/550°C was kept for 8 days. All material was transported and tens of micrometers thick single crystals of several millimeters of lateral size were obtained, however, rather unstable in the air avoiding almost any operation outside of the glovebox. The degradation process of these crystals was described in detail by Kratochvilova et al. [32].

In the second procedure, the work of Son et. al [3] was reproduced. First, the material was gradually heated up to 400°C for four days and then a gradient of 400/320°C was kept for 2 weeks. The process was significantly slower than the first one, approximately half of the input material remained untransported in the cold part. Nevertheless, we have obtained very large single crystals (several of them exceeding 0.5 $cm^2$) of a very reflective surface. Single crystals prepared by this procedure are much more stable in the air, for 2 hours without any surface degradation effect. Traces of degradation began to appear only locally in places where the single crystal was mechanically deformed, however, the degradation remained on the surface and provoked a negligible bulk effect after 5 hours. The chemical composition of single crystals was verified by scanning electron microscopy (SEM) using a Tescan Mira I LMH system equipped with an energy-dispersive X-ray detector (EDX) Bruker AXS. The EDX analysis revealed a single-phase single crystal of the stoichiometric $VI_3$ composition.

*Optical Experiments*. For Raman studies of single crystals, a Renishaw RM1000 Micro-Raman spectrometer with Bragg filters was used, equipped with an Oxford Instruments Microstat continuous flow optical He cryostat. The measurements were performed using an $Ar^+$ ion laser operating at 514.5 nm in the backscattering geometry. Low-temperature infrared (IR) transmittance measurements in the frequency range 20 – 660 $cm^{-1}$ (0.6-20 THz) were performed using a Bruker IFS-113v Fourier-transform IR spectrometer equipped with a liquid-He-cooled Si bolometer (1.6 K) serving as a detector. For the low-temperature IR and THz spectroscopies, continuous-He-flow cryostats (Optistat, Oxford Instruments) with polyethylene and mylar windows, respectively, were used. Temperature-dependent THz spectra of complex transmittance between 0.5 and 3 THz (16-100 $cm^{-1}$) were obtained using a custom-made time-domain spectrometer utilizing a Ti: sapphire femtosecond oscillator [60].

*Theoretical calculations*. Phonon dynamical matrix elements were calculated self-consistently using density functional theory (DFT) calculations in the band structure program ELK [61] for each phonon wave vector $k$ inside a supercell commensurate with $k$. These calculations employed the full-potential linear augmented plane wave (FPLAPW) method combined with the generalized gradient approximation (GGA) parametrized by Perdew-Burke- Ernzerhof [62]. This exchange-correlation functional represents an appropriate choice for phonon studies. Spin-orbit coupling (SOC) is known to play a key role in this system [12, 48] and has been included in the calculation. Hubbard U allows us to describe approximatively the strong electron correlations present in the system and form the correct band gap in the calculated electronic structure, while standard LDA calculation would lead to a metallic solution. Our calculations have used U = 3.8 eV added to the Hamiltonian within the fully localized limit double counting treatment [63] and a band gap of size around 0.6 eV is predicted. The full Brillouin zone has been sampled by about 800 $k$-points. Increased accuracy of expansion into spherical harmonics has been used with $l_{max}$ = 14

ACKNOWLEDGMENT

This work is a part of the research project GAČR 21-06083S which is financed by the Czech Science Foundation. The work of D.R. was supported by the Grant Agency of the Czech Technical University in Prague (Project No. SGS22/182/OHK4/3T/14) The $VI_3$ single crystals have been grown and characterized in the Materials Growth and Measurement Laboratory MGML (see: http://mgml.eu) which is supported by the program of Czech Research Infrastructures (project no. LM2018096). This work was also supported by the Grant No. CZ.02.1.01/0.0/0.0/15_003/0000487 (OP VVV project MATFUN). The authors are indebted to Dr. Martin Veis for kindly providing information on Raman data recorded on $VI_3$ in the magnetic field. The authors are indebted to Dr. Ross Colman for critical reading and correcting the manuscript.

REFERENCES

(1) McGuire, M. A. Crystal and Magnetic Structures in Layered, Transition Metal Dihalides and Trihalides. *Crystals* **2017**, *7* (5), 121.

(2) McGuire, M. A.; Dixit, H.; Cooper, V. R.; Sales, B. C. Coupling of Crystal Structure and Magnetism in the Layered, Ferromagnetic Insulator CrI3. *Chemistry of Materials* **2015**, *27* (2), 612-620. DOI: 10.1021/cm504242t.

(3) Son, S.; Coak, M. J.; Lee, N.; Kim, J.; Kim, T. Y.; Hamidov, H.; Cho, H.; Liu, C.; Jarvis, D. M.; Brown, P. A. C.; et al. Bulk properties of the van der Waals hard ferromagnet VI$_3$. *Physical Review B* **2019**, *99* (4), 041402. DOI: 10.1103/PhysRevB.99.041402.

(4) Kong, T.; Stolze, K.; Timmons, E. I.; Tao, J.; Ni, D.; Guo, S.; Yang, Z.; Prozorov, R.; Cava, R. J. VI3—a New Layered Ferromagnetic Semiconductor. *Advanced Materials* **2019**, *31* (17), 1808074. DOI: https://doi.org/10.1002/adma.201808074.

(5) Tian, S.; Zhang, J.-F.; Li, C.; Ying, T.; Li, S.; Zhang, X.; Liu, K.; Lei, H. Ferromagnetic van der Waals Crystal VI3. *Journal of the American Chemical Society* **2019**, *141* (13), 5326-5333. DOI: 10.1021/jacs.8b13584.

(6) Gati, E.; Inagaki, Y.; Kong, T.; Cava, R. J.; Furukawa, Y.; Canfield, P. C.; Bud'ko, S. L. Multiple ferromagnetic transitions and structural distortion in the van der Waals ferromagnet VI$_3$ at ambient and finite pressures. *Physical Review B* **2019**, *100* (9), 094408. DOI: 10.1103/PhysRevB.100.094408.

(7) Doležal, P.; Kratochvílová, M.; Holý, V.; Čermák, P.; Sechovský, V.; Dušek, M.; Míšek, M.; Chakraborty, T.; Noda, Y.; Son, S.; et al. Crystal structures and phase transitions of the van der Waals ferromagnet VI$_3$. *Physical Review Materials* **2019**, *3* (12), 121401. DOI: 10.1103/PhysRevMaterials.3.121401.

(8) Marchandier, T.; Dubouis, N.; Fauth, F.; Avdeev, M.; Grimaud, A.; Tarascon, J.-M.; Rousse, G. Crystallographic and magnetic structures of the VI3 and LiVI3 van der Waals compounds. *Physical Review B* **2021**, *104* (1), 014105. DOI: 10.1103/PhysRevB.104.014105.

(9) Hao, Y.; Gu, Y.; Gu, Y.; Feng, E.; Cao, H.; Chi, S.; Wu, H.; Zhao, J. Magnetic Order and Its Interplay with Structure Phase Transition in van der Waals Ferromagnet VI$_3$. *Chinese Physics Letters* **2021**, *38*, 096101. DOI: 10.1088/0256-307x/38/9/096101.

(10) Yan, J.; Luo, X.; Chen, F. C.; Gao, J. J.; Jiang, Z. Z.; Zhao, G. C.; Sun, Y.; Lv, H. Y.; Tian, S. J.; Yin, Q. W.; et al. Anisotropic magnetic entropy change in the hard ferromagnetic semiconductor VI$_3$. *Physical Review B* **2019**, *100* (9), 094402. DOI: 10.1103/PhysRevB.100.094402.

(11) Koriki, A.; Míšek, M.; Pospíšil, J.; Kratochvílová, M.; Carva, K.; Prokleška, J.; Doležal, P.; Kaštil, J.; Son, S.; Park, J. G.; et al. Magnetic anisotropy in the van der Waals ferromagnet VI$_3$. *Physical Review B* **2021**, *103* (17), 174401. DOI: 10.1103/PhysRevB.103.174401.

(12) Sandratskii, L. M.; Carva, K. Interplay of spin magnetism, orbital magnetism, and atomic structure in layered van der Waals ferromagnet **VI$_3$**. *Physical Review B* **2021**, *103* (21), 214451. DOI: 10.1103/PhysRevB.103.214451.

(13) Kim, K.; Lee, J. U.; Cheong, H. Raman spectroscopy of two-dimensional magnetic van der Waals materials. *Nanotechnology* **2019**, *30* (45), 10, Review. DOI: 10.1088/1361-6528/ab37a4.

(14) Zhang, Y.; Wu, X.; Lyu, B.; Wu, M.; Zhao, S.; Chen, J.; Jia, M.; Zhang, C.; Wang, L.; Wang, X.; et al. Magnetic Order-Induced Polarization Anomaly of Raman Scattering in 2D Magnet CrI3. *Nano Letters* **2020**, *20* (1), 729-734. DOI: 10.1021/acs.nanolett.9b04634.

(15) Lei, M.; Coh, S. Large cross-polarized Raman signal in CrI$_3$: A first-principles study. *Physical Review Materials* **2021**, *5* (2), 025202. DOI: 10.1103/PhysRevMaterials.5.025202.

(16) Tian, Y.; Gray, M. J.; Ji, H.; Cava, R. J.; Burch, K. S. Magneto-elastic coupling in a potential ferromagnetic 2D atomic crystal. *2D Materials* **2016**, *3* (2), 025035. DOI: 10.1088/2053-1583/3/2/025035.

(17) Milosavljević, A.; Šolajić, A.; Djurdjić-Mijin, S.; Pešić, J.; Višić, B.; Liu, Y.; Petrovic, C.; Lazarević, N.; Popović, Z. V. Lattice dynamics and phase transitions in Fe$_{3-x}$GeTe$_2$. *Physical Review B* **2019**, *99* (21), 214304. DOI: 10.1103/PhysRevB.99.214304.
11

**Supplementary Information for**

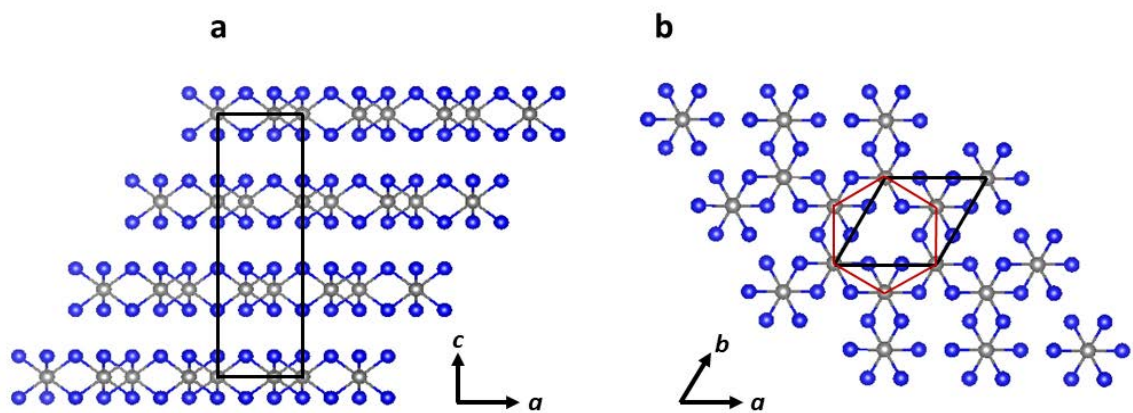

**Figure S1| The crystal structure of the VI₃ at room temperature. a,** a-c plane projection of four V-I triple-layers. **b,** a-b plane projection of one V-I triple-layer. In silver color V atoms, in blue color Iodine atoms. The chosen hexagonal unit cell corresponds to work[1] where we shifted the unit cell along the *c*-axis and set the V atom to (0,0,0) position. The red hexagon emphasizes the in-plane honeycomb arrangement of V atoms.

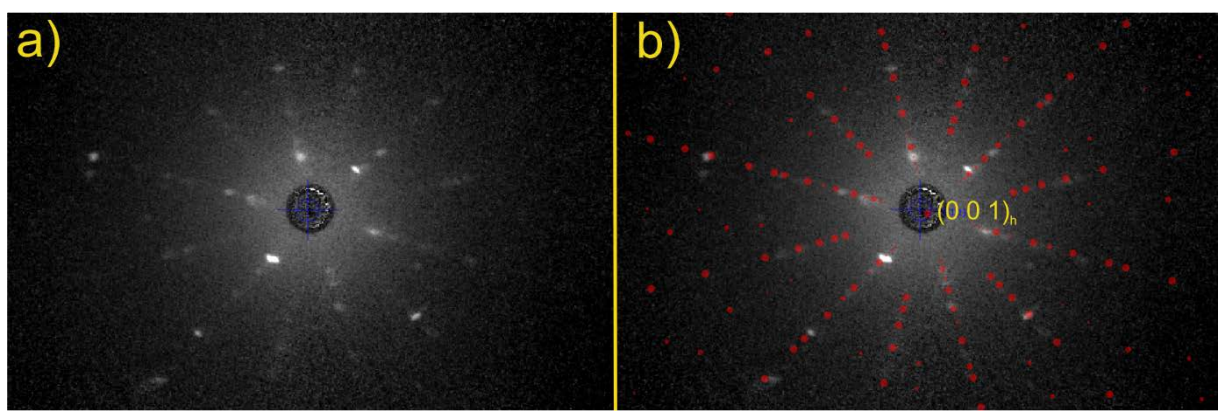

**Figure S2| Laue image of VI₃ single crystal.** a) bare image b) Comparison of measured pattern with the simulated one based on rhombohedral crystal structure model in work[1]. The Laue

diffraction was performed on VI3 sample placed on a glass bar and covered by blue tape to slow down the degradation process.

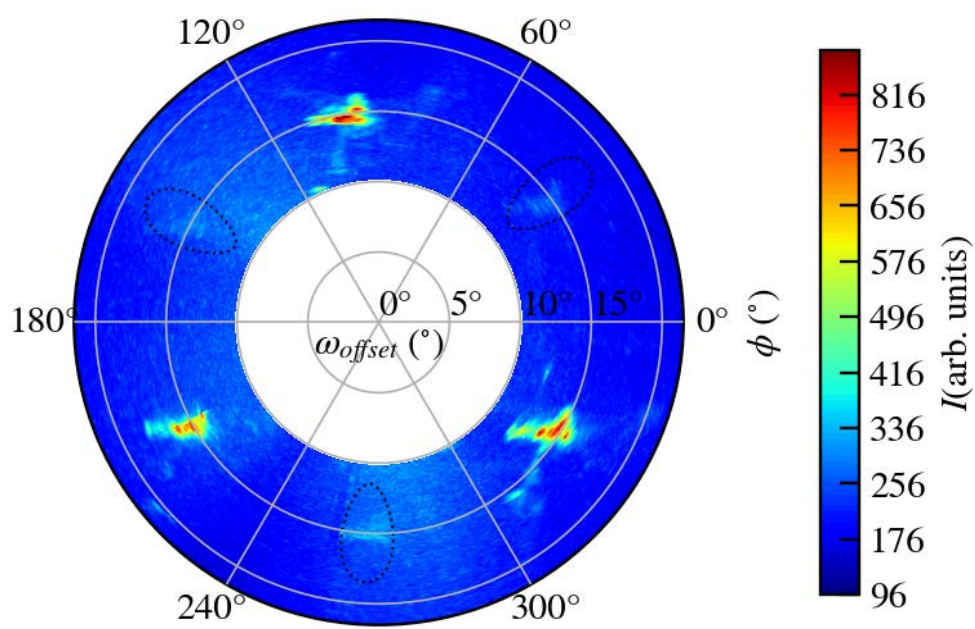

**Figure S3| Integrated intensity map.** Integrated intensity of HRM in $2\theta$ angle around (1 1 21), (-2 1 21), (1 -2 21) and (2 -1 21), (-1 2 21), (-1 -1 21) diffraction maxima. The 3-fold rotation symmetry is well visible as the (1 1 21), (-2 1 21), (1 -2 21) peaks are more intense than (2 -1 21), (-1 2 21), (-1 -1 21) (enclosed by dashed ellipse) matching the rhombohedral crystal structure model determined in work[1] at 250 K. All diffraction peaks should be separated by 120° in Φ direction. The deviation is caused by limited possibility for sample alignment in a powder diffractometer, which was equipped only by piezo-rotator in Φ direction. Sample on the piezo-rotator was rotated in Φ direction with 1° step. From the Figure it is also well visible the mosaicity of the single crystal, which is around 5°.

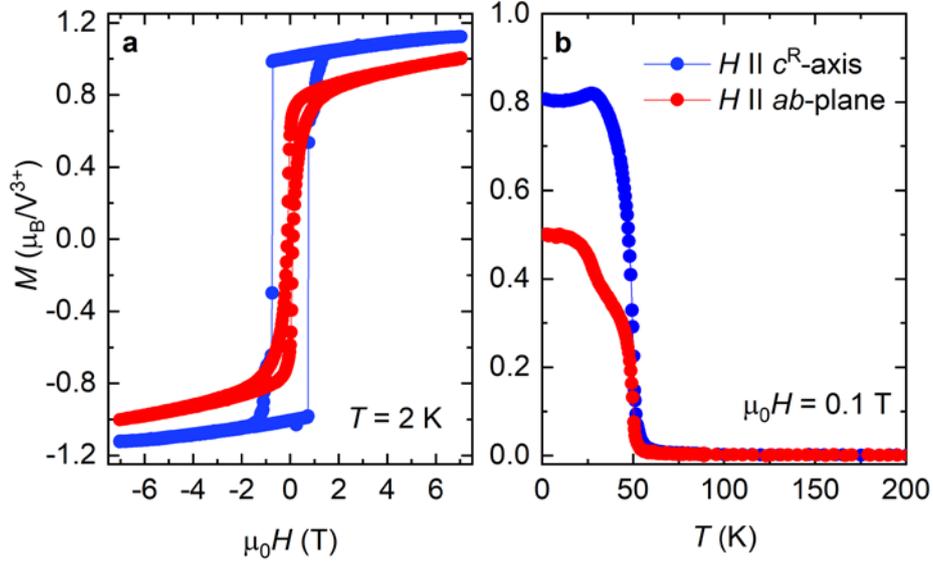

**Figure S4| Magnetization measurements of bulk VI₃. a,** magnetization isotherms measured at $T$ = 2 K. The magnetization in $\mu_0 H$ = 7 T reaches 1.12 $\mu_B$/V$^{3+}$ along $c^R$ – axis direction and 1 $\mu_B$/V$^{3+}$ along $ab$-plane direction. **b,** magnetization as a function of temperature measured after field cooling. At $T_C$ = 50 K a clear ferromagnetic transition is detected. At around 30 K the ab-plane magnetization projection shows a kink which is related to the order-order phase transition. In the same temperature interval, a small decrease of magnetization along the $c^R$ – axis is observed.

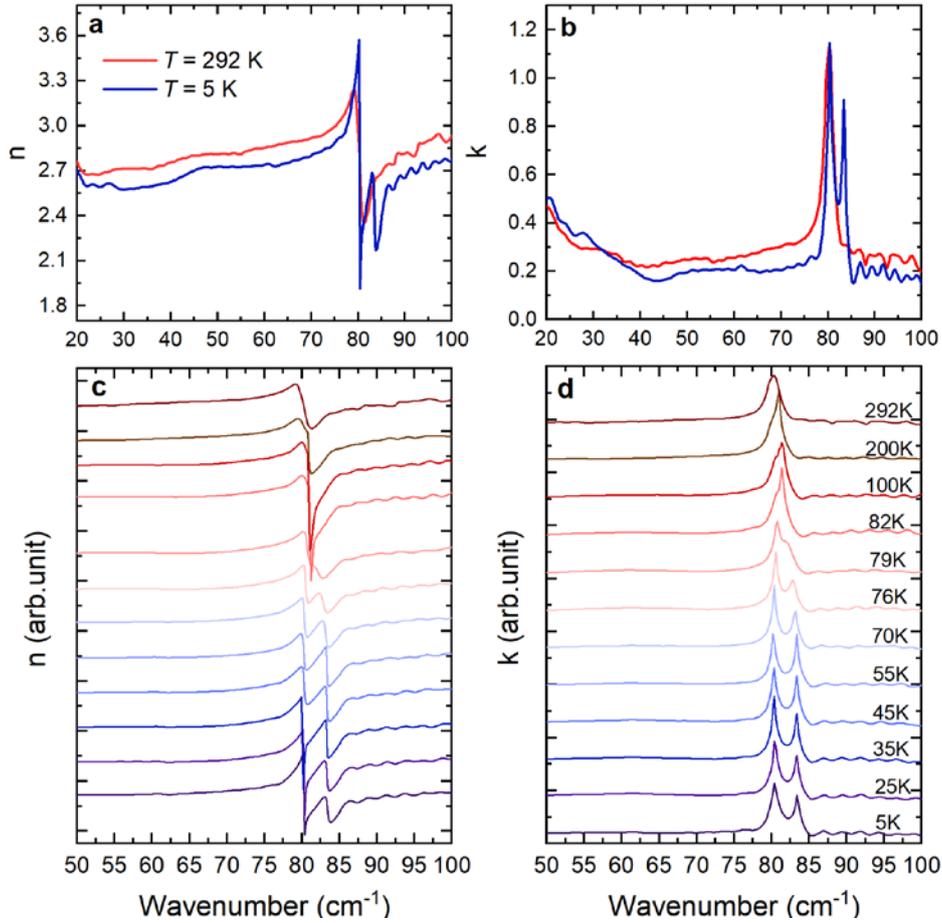

**Figure S5| THz spectra measured from room temperature to 5 K. a,** real part of the refraction index, n(ω). **b,** imaginary part of the refractive index, k(ω). We identified a phonon mode around 82 cm$^{-1}$ as an absorption peak and resonance dispersion in the spectral line shape of k(ω) and n(ω), respectively. **c,d,** temperature evolution n(ω) and k(ω). The splitting of the phonon mode is observed between 82 K and 79K confirming the lowering of the crystal symmetry.

Ungerade phonon modes assignments

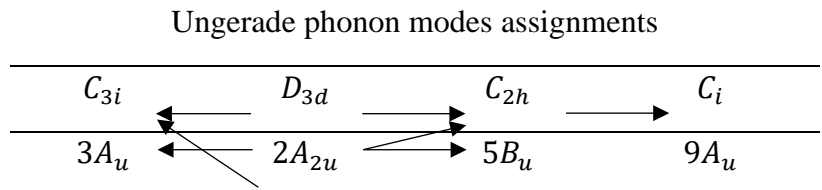

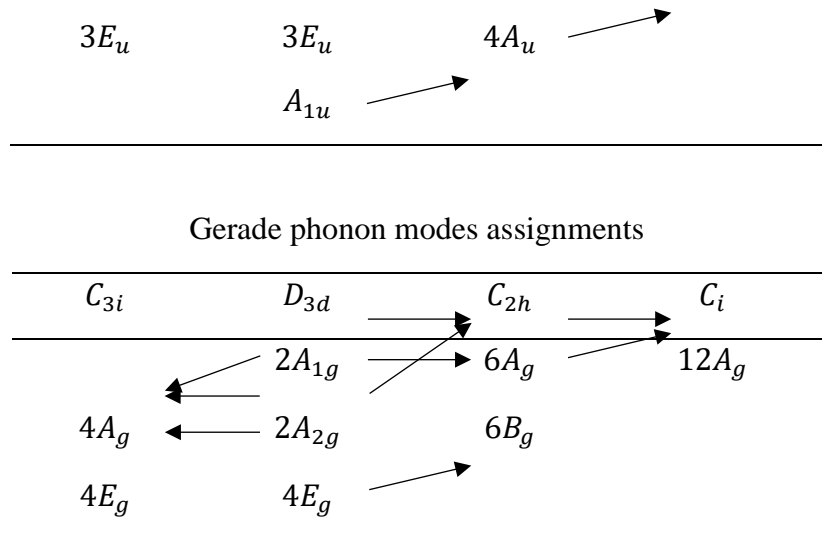

Gerade phonon modes assignments

| $C_{3i}$ | $D_{3d}$ | $C_{2h}$ | $C_i$ |
|---|---|---|---|
|  | $2A_{1g}$ | $6A_g$ | $12A_g$ |
| $4A_g$ | $2A_{2g}$ | $6B_g$ |  |
| $4E_g$ | $4E_g$ |  |  |

**Table S1| The relations between the irreducible representations of different point groups.** The point group $C_{3i}$ ($R\bar{3}$ space group) for trigonal phase and $C_{2h}$ ($C2/m$ space group) for the monoclinic phase do not obey relations group-subgroup. Therefore, no direct correspondence between the irreducible representations (irreps) of the groups exists. However, as the point group of the one triple-layer has higher symmetry which is $D_{3d}$ and is the same for both phases[2] (the difference between the structure is the triple-layer stacking sequence) we can find in-direct correspondence between the irreps via this point group. The point group $C_i$ for the trigonal phase is the subgroup of the $C_{2h}$ and therefore, the relations between the irreps can be easily find[3].

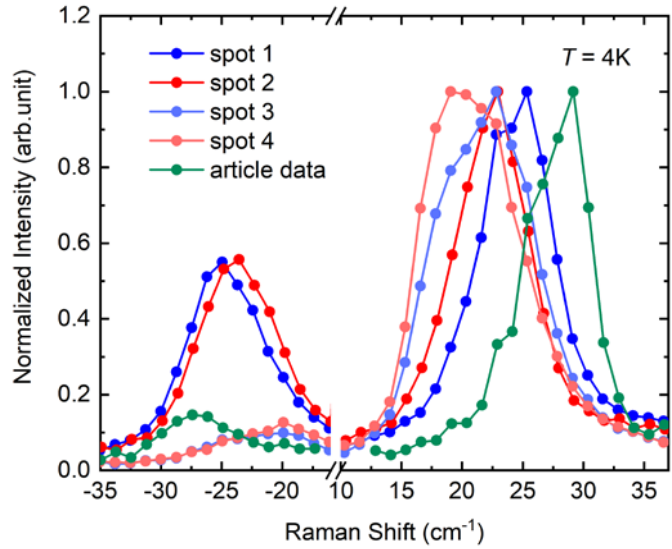

**Figure S6| Unpolarized Raman spectrum of the acoustic magnon.** We performed testing measurements of the Raman spectra in zero magnetic field which reveal small differences in the magnon frequency depending on the location of the laser spot. The frequency varies from 20 to 29 cm$^{-1}$. Two Raman spectra (spot 1 and spot 2) show relatively high intensity of the anti-Stokes line compared to the other spectra (each spectrum is normalized such that the magnon peak maximum of the Stokes line is at 1). This might indicate local heating of the sample by the laser, which could cause the frequency shift of the magnon mode. However, as shown in Figure 4b of the main paper, the heating of the sample decreases the magnon frequency which is not observed in our testing spectra. We add that for all spectra we used the same laser power.

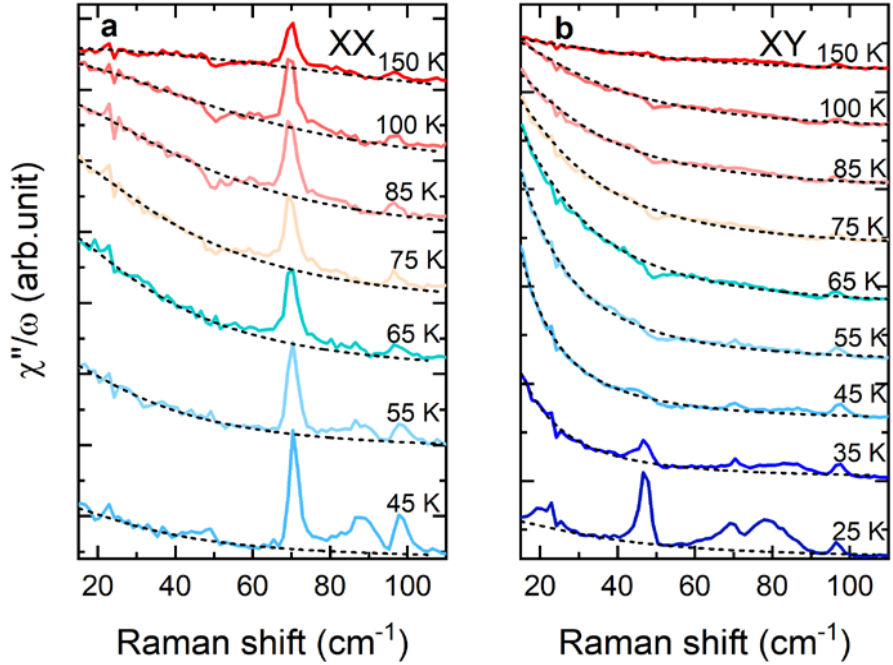

**Figure S7| temperature evolution of the Raman conductivity $\chi''/\omega$ measured in parallel (XX) and crossed (XY) polarization configurations. a,b,** The spectra in the temperature interval 150 K – 25 K (45 K for XX channel) were fitted with the Lorentzian function (black dashed lines) with the center of gravity at the Rayleigh line. The peaks and the Fano resonance superposed on the quasi-elastic signal were subtracted before the fitting procedure. To get the spectral weight we integrated the areas below the fits with the integration range of 12-100 cm$^{-1}$. One should keep in mind, that leaving out a part of the spectral weight (0 – 12 cm$^{-1}$) slightly decreases the accuracy of the finally established spectral weight values, especially at temperatures where the quasi-elastic scattering peak has small width. The small oscillation at 23 cm$^{-1}$, which occurs at each temperature in both polarization configuration, does not show any temperature evolution and is considered as an instrumental artifact.

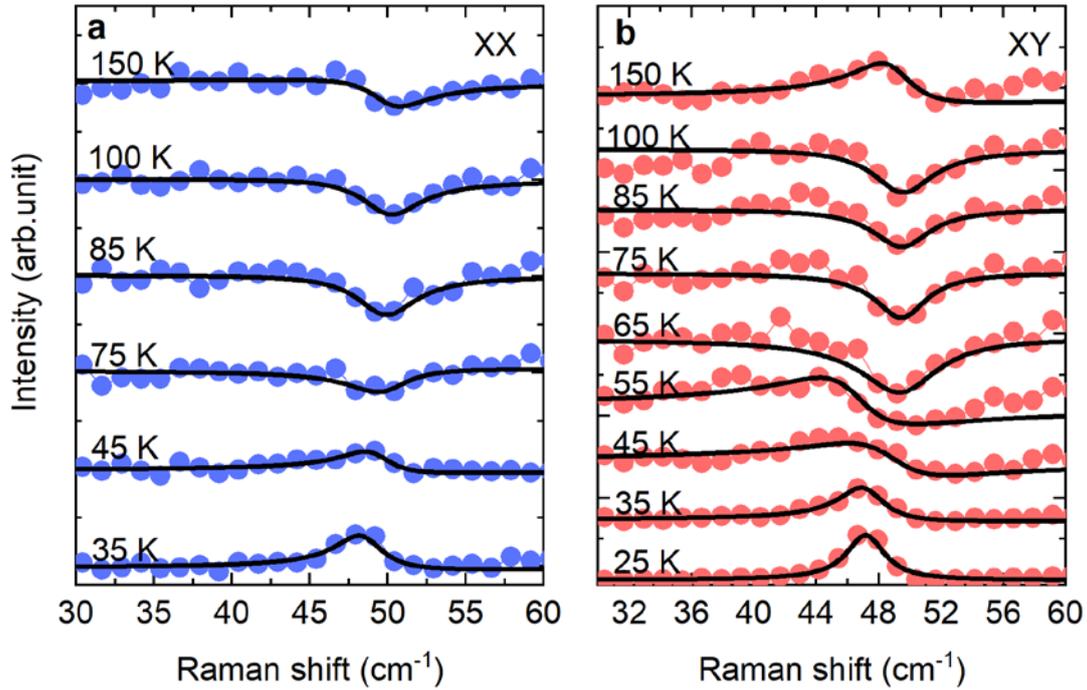

**Figure S8|** The Raman spectra of the $E_g^1$ phonon mode. **a,** for the parallel polarization configuration. **b,** for the crossed polarization configuration (the spectra in the temperature range 150 K - 45 K were multiplied by factor 5 and the spectrum at 35 K by factor 2 for better clarity). The spectra in both configurations were obtained by the subtraction of the quasi-elastic scattering contribution. The solid black lines represent the fits to Equation (9) in the main paper (Fano resonance line shape). We were not able to fit the Raman spectra taken at 55 K and 65 K for the XX polarization configuration by Equation (9) in the main paper as the $E_g^1$ phonon mode is unresolvable at these temperatures.